\documentclass[preprint,12pt]{elsarticle}

\usepackage{amssymb}

\journal{Physics Letters B}
\newcommand{\mev}{\, \mathrm{MeV}}

\newcommand{\hz}{\, \mathrm{Hz}}
\newcommand{\bra}{\langle}
\newcommand{\ket}{\rangle}

\begin{document}
\begin{frontmatter}



\title{Muon Capture on $^3$He and the Weak Structure of the Nucleon}


\author{Doron Gazit\corref{cor1}}

\address{Institute for Nuclear Theory, University of Washington, Box 351550, 98195 Seattle, Washington, USA}
\cortext[cor1]{ Corresponding author}
\ead{doron.gazit@mail.huji.ac.il}

\begin{abstract}
The weak form factors of the nucleon, including the induced pseudoscalar form factor 
and second class terms,  are constrained using a microscopic calculation of the weak capture process 
$^3\rm{He}(\mu^{-},\nu_\mu)^3\rm{H}$.  The calculation is parameter free, and yields a rate of $1499(16)\hz$, in agreement with the remarkable experimental measurement $1496(4)\hz$. The nuclear wave functions are obtained using the EIHH method with the Argonne $v_{18}$ nucleon--nucleon potential and the Urbana-IX three nucleon force. The weak currents in the nuclei are described using HB$\chi$PT formalism. The induced pseudoscalar form factor is found to agree with HB$\chi$PT prediction. The result is compatible with vanishing second class currents, with the tightest constraint to date on the conservation of vector current (CVC) hypothesis.
%
\end{abstract}


\end{frontmatter}
\begin{section}{Introduction}
The weak process in which a muon is captured by a nucleus provides an experimental hatch to various aspects of the 
fundamental forces. For decades, it has been used to constrain 
the properties and symmetries of the weak interaction, and to probe the
structure of the nucleus at relatively large momentum transfer 
$|\vec{q}|\sim m_\mu=105.6\mev$
\cite{Primakoff}. 

This sizable momentum transfer enhances the effect of terms proportional to $q$. One example is the induced-pseudoscalar form factor. This form factor has been the target of numerous studies in the past, which have revealed a contradiction between experimental and theoretical estimations. These ambiguities have induced a new experimental effort by the MuCap collaboration to measure the muon capture rate on protons ($\mu^-p$) \cite{MuCap}, which have resolved the contradiction. By now, the MuCap measurement \cite{MuCap}  has already reached a $\pm 2.4\%$ determination of the capture rate, and aims to $\pm 1\%$. This leads to the tightest experimental bound on the induced-pseudo scalar form factor of the nucleon, constraining its value to $\pm 15\%$ \cite{MuCap}.

Two additional poorly known form factors are the second class terms, in Weinberg's classification  \cite{Wein_SC}, who also assumed their vanishing. The vector second class term is also required to vanish by the conservation of vector current (CVC) hypothesis. These terms are demanded by Lorentz covariance, but change sign under G-parity transformation. However, this behavior only suppresses their value, due to the fact that isospin symmetry is only an approximate symmetry of the strong force.   The current experimental determination of both terms has not reached the sensitivity needed to test the theoretical estimations, and is still consistent with vanishing form-factors \cite{Standard_model_nuclear_tests_RMD}.

Further reduction of the uncertainties in these three form-factors demands sub-percentage experimental accuracy, which is hard to achieve in $\mu^-p$ process, since its rate is smaller than the free muon decay rate  by a factor bigger than 500, and since it results in the emission of neutral particles. 
These obstacles can be removed for heavier nuclei, as the capture rate is very sensitive to the
nuclear charge, scaling as $Z^4$.
Alas, due to the strong correlation between nucleons, theoretical 
microscopic studies with sub-percentage precision, are possible only in very light nuclei. 

However, one capture process fits the theoretical limitations, 
and has been measured to a very high accuracy. The capture  
of muon on $^3\rm{He}$ which results in a triton,
\begin{equation} \label{eq:process}
\mu^{-} + ^{3}\rm{He} \rightarrow \nu_\mu + ^{3}\rm{H},
\end{equation}
with a measured capture rate of 
$\Gamma(\mu^{-} + ^{3}\rm{He} \rightarrow \nu_\mu + ^{3}\rm{H})^{exp}=1496(4) \hz$, 
i.e. a $\pm 0.3\%$ precision \cite{3He_mu_meas}. 
This precision measurement has already induced a number of theoretical works \cite{Gorringe2003}, 
among them are also microscopic theoretical studies of the reaction 
\cite{Gorringe2003,Truhlik_MU,Marcucci_MU}. 
The main conclusion of these works is that theoretical evaluation should include both state of the art 
description of the nuclear states, and a correct description of the weak interaction between 
the muon and the nucleus, including the interaction with meson exchange currents (MEC) in the nucleus. 
Though these studies have achieved very high accuracy in the nuclear sector, 
they missed an important, recently discovered, ingredient. Lately \cite{MUCAP_RC}, the electroweak radiative 
corrections to the capture process have been calculated. The results indicate that the 
enhancement factor due to this effect is $\rm{RC(He)}=0.030(4)$. 
The authors of Ref.~\cite{MUCAP_RC} demonstrate that this destroys the good agreement 
of the theoretical study of reaction (\ref{eq:process}), done in Ref.~\cite{Marcucci_MU}. In light 
of this development, a new theoretical evaluation of the process is called for.

In the current work we cope with this challenge. We study theoretically reaction (\ref{eq:process}), and use it to put constraints on the weak form factors of the nucleon, including the second class terms, which was not done previously in microscopic calculations. We use a hybrid approach, which has been proved efficient in describing different weak processes with $A=2-4$ nuclei \cite{PA03,PhD}, using phenomenological Hamiltonian for the nuclear states, and heavy baryon chiral perturbation theory to describe the weak interaction of the muon with the nucleus. The latter puts on common ground the single nucleon current and meson exchange currents, a fact which increases the reliability of the calculation. In addition, the calculation is parameter free, thus can be used to make predictions.  

The letter is built as follows. In the next section we outline shortly the standard model formalism for calculating the capture process. The solution
of the nuclear problem is described in Sec.~3, followed by the derivation of the weak currents in the nucleus in Sec.~4. In Sec.~5 we give the theoretical capture rate and present an error estimation on the results. We discuss the consequences of the results on the weak form factors in the last section.

\end{section}
\begin{section}{Theoretical Formalism}
We start with a brief reminder of muon capture process, and the formalism used in the calculation. 
The muonic atom, a bound state of a muon and a nucleus, is unstable. It has two main possible decay schemes, either through
free muon decay to lighter leptons, or through a muon capture by the nucleus. 
The capture is a weak process, where the negative muon 
interacts with the nucleus through the exchange of heavy W$^{-}$ boson.
As the momentum transfer in the process is much smaller than the mass of the
$W^{-}$ boson, the weak interaction Hamiltonian is given by
 $\hat{H}_{W}=-\frac{G|{V_{ud}}|}{\sqrt{2}}\int {d^{3}x
\hat{j}^{+}_{\mu }(\vec{x}) \hat{J}^{-\mu }(\vec{x})}$, where $G=1.166371(6) 
\times 10^{-11} \mev^{-2}$ is the Fermi coupling constant \cite{PDBook}, 
$V_{ud}=0.9738(4)$ is the CKM matrix element mixing $u$ and $d$ 
quarks involved in the process \cite{PDBook}, $\hat{j}^{+}_{\mu }(\vec{x})$ 
is the lepton charge raising current, and $\hat{J}^{-\mu }$ 
is the nuclear charge lowering current. 

It is straight forward to evaluate the lepton current. We first note that since $Z\alpha \ll 1$
($Z$ is the nuclear charge, and $\alpha$ is the fine structure constant)
the muon bound in the atom can be regarded as non-relativistic
. Moreover, the initial 
(muonic) atom has a Bohr radius much larger than the nucleus radius, thus penetration operators 
are negligible for the needed accuracy \cite{Wick_private}. The discussion
in Ref.~\cite{MU_Beyond_SM} shows that in these conditions, one can approximate 
the lepton current as a current of point like Dirac particles, whose states are described by plane
waves, multiplied by a correction factor. This factor takes into account the initial bound state of the 
muon in the atom, and the charge distribution of the nucleus. For $^3$He, this factor is 
calculated in Ref.~\cite{WALECKA:BOOK,MU_Beyond_SM}: $|\psi_{1s}^{av}|^2 = \mathcal{R}\frac
{(Z\alpha M_r)^3}{\pi}$, where $\mathcal{R}=0.979$, and $M_r=(M_{^3\rm{He}}^{-1} + m_\mu^{-1})^{-1}$
is the reduced mass of the muonic atom.

We use the Golden rule to write the capture rate \cite{WALECKA:BOOK,Marcucci_MU}:
\begin{equation} \label{eq:rate}
\Gamma ={\frac{{2 G^2 |V_{ud}|^2 E_\nu^2}}{2J_{{\rm{^3 He}}}+1} 
\left( 1- \frac{E_\nu}{M_{^3{\rm{H}}}}\right)} |\psi_{1s}^{av}|^2 {\Gamma_N},
\end{equation}
where $J_{\rm{^3 He}}=\frac{1}{2}$ is the total angular momentum of the ${\rm{^3 He}}$. The effects
of the nuclear interaction are embedded in the nuclear matrix element ${\Gamma_N}$, which can be 
written using multipole decomposition:
\begin{eqnarray} \label{eq:NME}
\nonumber \Gamma_N & = &  {\sum_{J=0}^\infty \left| \bra ^3{\rm{H}} 
\| \hat{C}_J - \hat{L}_J \| ^3{\rm{He}} \ket \right|^2 
 + }\\ &+&\sum_{J=1}^\infty \left| \bra ^3{\rm{H}} \| \hat{E}_J 
- \hat{M}_J \| ^3{\rm{He}} \ket \right|^2 .
\end{eqnarray}

$\hat{C}_{J}, \hat{L}_{J}, \hat{E}_{J}, \hat{M}_{J}$ are the Coulomb,
longitudinal, transverse electric and transverse magnetic multipole operators
of angular momentum $J$, built from the charged nuclear current. One should 
notice that since the total angular momentum of the ${\rm{^3 He}}$ and  ${\rm{^3 H}}$ is 
$\frac{1}{2}$, and both posses positive parity, only some multipoles survive:
$C_0^V,\,L_0^V,\,C_1^A,\,L_1^A,\,E_1^A,\,M_1^V$. The superscript 
 $A$ ($V$) stands for operators of axial (vector) symmetry. As the $\chi$PT vector current
satisfies CVC, the vector Coulomb and Longitudinal reduced matrix elements are related: 
$<L^V_{J}> = -\frac{\omega}{|\vec{q}|}<C^V_J>$.

From this discussion, it is clear that the needed information 
is the structure of the weak currents in the nucleus, 
and the wave functions of the $^3$He and triton. We will discuss these issues in the
following two sections.
\end{section}
\begin{section}{The Nuclear Wave Functions}
The evaluation of $\Gamma_N$ in Eq.~(\ref{eq:rate}), demands 
the solution of the three--body nuclear problem, for the ground states
of the triton and  ${\rm{^3 He}}$. We solve the Schr\"{o}dinger equation 
microscopically using the effective
interaction in the hyperspherical harmonics (EIHH) approach
\cite{EIHH}, as implemented in the {\sc{nbody}} fortran code\cite{nbody}. 
In a previous study of reaction~(\ref{eq:process}),
Marcucci {\it{et al.}} \cite{Marcucci_MU} have shown that the capture is essentially
independent of the nuclear force, as long as it describes correctly the binding 
energies of the trinuclei. Thus, the nuclear Hamiltonian is taken as the
nucleon-nucleon potential Argonne $v_{18}$ (AV18) \cite{AV18} with
the Urbana IX (UIX) \cite{UIX} three nucleon force. This
Hamiltonian has been used successfully to reproduce the spectra of the trinuclei
as well as other light nuclei \cite{UIX}, and also electro-weak reactions with light
nuclei \cite{GA06,MA01,pp_fusion}. Table~\ref{tab:trinuclei} shows a comparison of our numerical results 
for the binding energies of the trinuclei with the experimental measurements. Also shown is a comparison
to the calculation made using two other {\it{ab-initio}} methods,
solving the Fadeev-Yakubovski (FY) equations, and using the Correlated Hyperspherical Harmonics (CHH) method\cite{Nogga}. 

\begin{table}
\begin{center}
\begin{tabular}{|c||c|c|}
\hline
 & \multicolumn{2}{c|} {\bfseries Binding Energy [$\mev$]} \\ \cline{2-3}
 \raisebox{1.5ex}[0pt]{\bfseries Method}   &     $^3$H      &      {$^3$He}      \\\hline \hline
 EIHH     & $8.471(2)$ & $7.738(2)$     \\
 CHH      & $8.474   $ & $7.742   $     \\
 FY       & $8.470   $ & $7.738   $      \\\hline
 Experiment & $8.482   $ & $7.718   $      \\\hline
\end{tabular}
\end{center}
\caption{Binding energies of $^3$H and $^3$He calculated using
AV18+UIX Hamiltonian model compared to the same calculation done by
using FY equations and CHH method \cite{Nogga}. For the EIHH calculation,
the number in parenthesis indicates the numerical error. Also shown
are the experimental values.} \label{tab:trinuclei}
\end{table}

\end{section}
\begin{section}{Weak Currents in The Nucleus}
The main difference between previous works and the current one, is in the details of the
nuclear current. The formal structure of the nuclear charge lowering current is dictated 
by the Standard Model: $\hat{J}_{\mu }^{-}=\frac{\tau _{-}}{2} \left(\hat{J}^V_{\mu
}+\hat{J}^A_{\mu}\right)$, where the superscript 
 $A$ ($V$) stands for current with axial-vector (vector) symmetry. 
$\tau_{-}$ is the isospin lowering operator. The axial and vector currents
are more complicated, as they are affected from the strong interaction, which governs
the dynamics in the nucleus. To the best of our knowledge the fundamental theory 
of the strong interaction is QCD. Thus, in principle the currents should be 
extracted from the QCD lagrangian. This is, however, impossible due to the non-
perturbative character of QCD at low energy. 

A possible solution to this problem is found in an effective field theory (EFT) approach 
to QCD \cite{WE90}, that is $\chi$PT. In $\chi$PT, one uses
the fact that the QCD lagrangian is chirally symmetric in the limit of massless 
up and down quarks, i.e. it is invariant under global $SU(2)_L\times SU(2)_R$ transformations.
The absence of parity doublets in the low mass hadron spectrum shows that
the axial symmetry is spontaneously broken at low energies. The pions are identified as the Goldstone
bosons of the chiral symmetry breaking, and their mass is interpreted as a result 
of the fact that the $u$ and $d$ quarks have mass, albeit small. Thus, $\chi$PT constructs 
a low energy lagrangian, that consists of nucleons and pions, and posses the symmetries of
QCD. Furthermore, Weinberg \cite{WE90} has given a recipe for organizing this
lagrangian in terms of $(Q/\Lambda)^\nu$, where $Q$ is the typical momentum in the process (about
$100 \mev$ for muon capture),
or the pion mass, $\Lambda$ is of the order of the EFT breakdown scale, and $\nu \ge 0$. 
An additional simplification is due to the large nucleon mass, which is of the order of the
chiral symmetry breaking scale, which allows non-relativistic expansion of the lagrangian,
the so called heavy baryon $\chi$PT (HB$\chi$PT). The nuclear currents, from this point of view, are
N\"{o}ther currents derived from the axial and vector symmetries of this lagrangian. 
In the last two decades, a huge amount of work has been done to derive the nuclear currents 
from the $\chi$PT lagrangian. 

The nuclear currents are derived in Ref.~\cite{PA03,PhD}, from a $\chi$PT lagrangian. 
The currents are expanded to next--to--next--to--next--to--leading order. 
As expected, one finds a nucleonic current, i.e. the impulse approximation, and meson exchange
currents (MEC). The nucleonic current achieved in this formalism is
identical in its form to the usual impulse approximation (IA). Its vector part
takes the form:
\begin{equation} \label{eq:V_IA}
\small{\hat{J}^V_{\mu}({\mathrm{IA}})={\bar{u}(p') \left[ F_V(q^2) \gamma^\mu 
+ \frac{i}{2M_N}F_M(q^2) \sigma^{\mu\nu}q_\nu \right] u(p)}},
\end{equation}
whereas the axial part is
\begin{equation}\label{eq:A_IA}
\hat{J}^A_{\mu}({\mathrm{IA}}) = - {\bar{u}(p') \left[ G_A(q^2) \gamma^\mu \gamma_5
+ \frac{G_P(q^2)}{m_\mu} \gamma_5 q_\mu \right] u(p)}.
\end{equation}
Here, $M_N$ is the nucleon mass, $m_\mu$ is the muon mass, and $u(p)$ is 
the Dirac spinor of the nucleon of momentum $p$. In order to keep with the 
power counting of HB$\chi$PT, we expand Eq.~(\ref{eq:V_IA}-\ref{eq:A_IA})
in powers of $1/M_N$, up to $\mathcal{O}(M_N^{-3})$. The currents contain four form factors. 
$F_V$ and $G_A$ are the vector and axial form factors, $F_M$ is the weak magnetism
form factor, and $G_P$ is the induced pseudo-scalar form factor. In this order
of $\chi$PT, the form factors contain one-pion-loop correction reflected in
their $q^2$ dependence. The first three form-factors are very well determined 
experimentally. The vector and weak magnetism are just isospin rotations of the
electro-magnetic form factors, $F_V(0)=1$ and $F_M(0)=3.706$, and are extrapolated
to the kinematics of reaction~(\ref{eq:process}), i.e. to $q^2=-0.954 m_\mu^2$,
$F_V=0.974(1)$ and $F_M=3.580(3)$. 
The momentum dependence of the axial form factor is $G_A(q^2)=g_A (1+\frac{r_A^2}{6}q^2)$, with $g_A=1.2695(29)$
and the axial radius of the nucleon $r_A^2=0.43(3)\, \rm{fm}^2$, thus $G_A(-0.954m_\mu^2)=1.245(4)$.

For the induced pseudoscalar coupling, HB$\chi$PT prediction to one loop corrections \cite{Kaiser} coincides 
with the well known Adler-Dothan formula:
\begin{equation} \label{eq:G_P}
G_P(q^2)=\frac{2m_\mu g_{\pi p n} f_\pi}{m_\pi^2-q^2}-\frac{1}{3}g_A m_\mu M_N r_A^2
\end{equation}
with $g_{\pi p n} = 13.05(20)$ and $f_\pi=92.4(4) \mev$. In our case
$g_p=7.99(20)$.

The single nucleon currents are invariant under G-parity transformations\footnote{The 
combination of charge conjugation and a rotation in isospin space.}, and were thus 
classified by Weinberg \cite{Wein_SC} as first class currents. In principle, the electro-weak theory
does not exclude the possibility of second class currents, which change sign under these transformations.
It is clear that G-parity breaking currents can rise from the fact that isospin symmetry is 
only approximate, i.e. of the order of $\frac{|m_u-m_d|}{M_N}$. Using general symmetry arguments, 
their contribution to the single nucleon currents of Eq.~(\ref{eq:V_IA}-\ref{eq:A_IA}) can be written 
as $\delta\hat{J}^V_{\mu}({\mathrm{IA}})= \frac{g_s}{m_\mu}q^\mu$, and 
$\delta\hat{J}^A_{\mu}({\mathrm{IA}})=-i \frac{g_t}{2M_N}\sigma ^ {\mu\nu}
q_\nu \gamma_5$. In addition, the term added to the vector current breaks the well 
known conservation of vector current (CVC) hypothesis. We will test the constraints reaction~(\ref{eq:process}) 
can put on these currents in the discussion section of the paper.

As stressed above, incorporation of MEC is essential for a percentage level 
prediction of the capture rate. In the HB$\chi$PT formalism MEC appears at $\nu=2$ \cite{PA03}. 
In configuration space, the MEC are Fourier transform of propagators with a cutoff
$\Lambda$. This leads to a cutoff dependence, which is renormalized by a 
cutoff dependent counterterm. Due to the limited scope of this letter, we refer the reader to Ref.~\cite{PA03,PhD}, 
for the explicit form of the MEC operators. For the muon capture process, all 
low-energy coefficients in the MEC can be determined from pion--nucleon 
scattering, except for one counterterm $\hat{d}_r(\Lambda)$, which characterizes the strength of
a two--nucleon contact term. 

We fix $\hat{d}_r(\Lambda)$ by reproducing the experimental triton half--life of $12.264 \pm 0.018$ years, 
which corresponds to $E_1^A$ strength of $0.6835\pm 0.001$ \cite{pp_fusion,rus_trit}\footnote{This value is slightly different than the one used by Ref.~\cite{PA03,pp_fusion},
consequently changing $\hat{d}_r(\Lambda)$ calibration.
}. Due to the fact that 
the value used here for the triton half life is used for the first time, we use a conservative 
error estimation due to it, multiplying by a factor of $2$ the quoted experimental error bar (i.e. we use $0.002$). The resulting cutoff dependence of $\hat{d}_r(\Lambda)$ is:
\begin{eqnarray} \label{eq:d_r}
\nonumber
\hat{d}_r(\Lambda=500\mev) & = & 1.05(6)_t  (0)_{N} \\
\hat{d}_r(\Lambda=600\mev) & = & 1.82(7)_t  (1)_{N} \\ \nonumber
\hat{d}_r(\Lambda=800\mev) & = & 3.88(9)_t  (2)_{N} 
\end{eqnarray}  
The first error is due to the triton half life, while the second is due to 
numerics.

This concludes the nuclear current needed for the calculation, and specifies the uncertainties 
in it.
\end{section}
\begin{section}{Theoretical Capture Rate and Error Estimation}
The results for the nuclear matrix element of the muon capture process, $\Gamma_N$, are listed in
table~\ref{tab:rates}. The results show a $9\%$ effect due to the MEC 
contribution, and a small effect due to the cutoff dependence of the 
HB$\chi$PT, within $0.3\%$. One can average this dependence, and arrive
at the prediction: $\Gamma_N=0.7075(10)$, when using 
the nominal values of the parameters throughout the letter. It is worthwhile noting that
the relative contribution of the MEC to this process is almost three times bigger than the 
MEC contribution to the triton half life. The extremely weak cutoff dependence shows that the
essential physics is captured in the HB$\chi$PT operators. 
 
\begin{table}
\begin{center}
\begin{tabular}{|c||c|c|c|c|}
\hline
&  & \multicolumn{3}{c|}  {\bfseries Total}\\ \cline{3-5}
  & \raisebox{1.5ex}[0pt] {\bfseries IA}   
  &     $\Lambda=500$      &      $\Lambda=600$ & $\Lambda=800$      \\\hline \hline
 $E_1^A$     & $0.5612$ & $0.5778$ & $0.5756$ & $0.5745$     \\
 $M_1^V$     & $0.1134$ & $0.1292$ & $0.1312$ & $0.1337$     \\
 $L_1^A$     & $0.2777$ & $0.2983$ & $0.3012$ & $0.2985$     \\
 $C_1^A$     & $0.0030$ & $0.0030$ & $0.0030$ & $0.0030$     \\
 $C_0^V$     & $0.3329$ & $0.3329$ & $0.3329$ & $0.3328$     \\ \hline
 $\Gamma_N$  & $0.6499$ & $0.7060$ & $0.7078$ & $0.7085$     \\ \hline \hline
\end{tabular}
\end{center}
\caption{The nuclear matrix element $\Gamma_N$ and its different multipole 
contributions, for the impulse approximation (IA) and the calculations which
include MEC (Total) for different cutoff values $\Lambda$ (in units of $\mev$).}
\label{tab:rates}
\end{table}

Thus, our final prediction for the capture rate is
\begin{equation} \label{eq:final}
\Gamma= 1499 (2)_\Lambda (3)_{\mathrm{NM}} (5)_t (6)_{\mathrm{RC}} \rm \hz,
\end{equation}
where the first error is due to the HB$\chi$PT cutoff, the second is due to
uncertainties in the extrapolation of the form factors to finite momentum
transfer, and in the choice of the specific nuclear model, 
the third error is related 
to the uncertainty in the triton half life, as reflected in Eq.~(\ref{eq:d_r}),
and the last error is due to theoretical uncertainty 
in the electroweak radiative corrections calculated for nuclei \cite{MUCAP_RC}.
This sums to a total error estimate of about $1\%$. 

The error estimation due to the choice of the specific nuclear model was discussed by 
Marcucci {\it{et.}} al.\cite{Marcucci_MU}. By considering different force models,
they found that using different nuclear potentials does not have a substantial effect 
on the capture rate, as long as the calculation reproduces the binding energies of the trinuclei. 
Their estimate for the uncertainty resulting from this was about 
$2\,\hz$. As the evaluation of HB$\chi$PT based potentials
evolves \cite{EFT}, one would be able to use a nuclear model of the same
microscopic origin as the currents, which could merge the nuclear model error and
the $\Lambda$ cutoff, and possibly reduce the estimated uncertainty. 

The experimental error in the triton half life, as mentioned earlier, is multiplied by $2$, which will
be reduced in the future, when the current measurement will become standard. 

The radiative corrections, which are the source of the largest contribution to the error estimation, are
not taken into account in previous studies \cite{Marcucci_MU,Truhlik_MU}. One could reduce the large 
uncertainty in this contribution, by extending the work in Ref.~\cite{MUCAP_RC} to include
higher order effects and incorporating nuclear effects. 

We sum the different error estimations linearly, since there might be correlations which we are not
aware of. This assumption seems wrong mainly for the radiative corrections, which seem independent of
the other contributions. Had we took a statistical sum for those, the error would have decreased to about $0.7\%$. However, a conservative estimation is appropriate due to the importance of the conclusions.
\end{section}
\begin{section}{Discussion}
The conservative error estimation still allows rather interesting conclusions. First, one notices that
the calculated capture rate agrees with the experimental measurement 
$\Gamma(\mu^{-} + ^{3}\rm{He} \rightarrow \nu_\mu + ^{3}\rm{H})^{exp}_{stat}=1496(4)\rm$$\hz$. We thus
conclude that the hybrid approach used in the current work, which is usually named EFT* \cite{PA03},
accurately predicts the capture rate. 

From this, one practical conclusion can be derived concerning the ability 
of the method of calculation to predict weak reaction rates in astrophysical environment, in particular in  
supernova environment. These reactions are usually unreachable experimentally, and their calculation
includes momentum transfer of few tens of $\mev$. The currents used in these calculations are based 
on extrapolation of $\beta$ decay surveys and theoretical consideration. By using the same currents 
to predict muon capture rates, the extrapolation becomes interpolation. This conclusion is also deduced by
Zinner {\it{et. al.}}\cite{Zinner}, who used RPA approach to calculate total muon capture rates in heavy nuclei.

However, the most interesting result concerns the weak form factors of the nucleon. In order to constrain the 
induced pseudoscalar and second class form factors, we take the following approach. In each case, we set all the other
form factors to their nominal value, and change this form factor in a way which keeps an overlap between the experimental rate 
and the theoretically allowed rate. The nominal value of the form factor is set to reproduce the experimental measurement.

The resulting constraint on the induced pseudoscalar form factor is:
\begin{equation}
g_P(q^2=-0.954 m_\mu^2) = 8.13 \pm 0.6 \,,
\end{equation}
in very good agreement with the HB$\chi$PT prediction of Eq.~(\ref{eq:G_P}). Together with the MuCap
results \cite{MuCap}, $g_P(q^2=-0.88 m_\mu^2) = 7.3 \pm 1.2 $, this is a great success to the HB$\chi$PT prediction.

A second conclusion concerns the contribution of second class currents. The axial G-parity breaking term was predicted, based on QCD sum-rules 
to be $\frac{g_t}{g_A}=-0.0152(53)$ \cite{Shiomi_SC}. Using this prediction does not change the 
result of the current calculation significantly (about $0.15\%$). Our constraint has a much larger error bar than this calculation, and agrees with
a vanishing form factor:
\begin{equation}
\frac{g_t}{g_A}  = -0.1   \pm 0.68\,.
\end{equation}
Wilkinson \cite{Wilkinson_SC} has collected the experimental data to get $|g_t|<0.3$ at $90\%$ 
CL, which is a factor of 2 better than the current limit. One has to still consider that the 
G-parity breaking terms can also excite mesonic currents in the nucleus, which were not taken
into account in the current discussion.

To date, the tightest constraints on the vector G-parity breaking term, $\delta\hat{J}^V_\mu$, 
related with the Conserved Vector Current (CVC) hypothesis, were made using a survey of 
superallowed $0^+ \rightarrow 0^+$ $\beta$ decays. This leads to a value of $g_s=0.01 \pm 0.27$ 
\cite{Standard_model_nuclear_tests_RMD}. Our calculation puts a much tighter limit, which can be considered experimental, 
on this form-factor:
\begin{equation}
g_s = -0.005 \pm 0.040,.
\end{equation}
Thus, the current limit provides the tightest constrain on $g_s$, and still agrees with CVC.

Summarizing, we have calculated the rate of the weak process $^3\rm{He}(\mu^{-},\nu_\mu)^3\rm{H}$. The calculation 
predicts a capture rate of $\Gamma=1499 \pm 16 \hz$, in accord with the measured rate $\Gamma=1496 \pm 4 \hz$. 
The error estimation has two main sources, uncertainties in the experimental triton half life, and in the calculation of 
radiative corrections to the process. The nuclear wave functions are calculated, using the EIHH method, with the phenomenological nuclear forces AV18+UIX. 
The hadronic currents within the nucleus are derived from HB$\chi$PT. Their low energy constants are constrained 
from low energy pion-nucleon scattering, and from the triton half--life. As a result, the calculation has no free-parameters. 
The induced pseudoscalar form factor is constrained to $\pm 8\%$, and agrees with HB$\chi$PT prediction\cite{Kaiser}. We show that this prediction is consistent with 
vanishing second class terms. The CVC hypothesis is confirmed to a new limit $|g_s|<0.045$. 
The calculation shows that nuclear {\it{ab initio}} calculations of muon 
capture process can have percentage level accuracy, and thus can be used as a quantitative test for the weak
structure of the nucleon and other properties of QCD at low energy.
\end{section}
\begin{section}{Acknowledgments}
The author acknowledges discussions with N. Barnea, N. T. Zinner, and W. C. Haxton. 
This work is supported by DOE grant number DE-FG02-00ER41132.
\end{section}

\end{document}